\documentclass[twocolumn,showpacs,preprintnumbers,amsmath,amssymb]{revtex4}

\usepackage{graphicx}
\usepackage{color}

\begin{document}

\title{Momentum distribution of a freely expanding Lieb-Liniger gas}

\author{D.~Juki\' c, B.~Klajn, and H.~Buljan}
\email{hbuljan@phy.hr}
\affiliation{Department of Physics, University of Zagreb, Bijeni\v cka c. 32, 10000 Zagreb, Croatia}

\date{\today}

\begin{abstract}
We numerically study free expansion of a few Lieb-Liniger bosons, which are 
initially in the ground state of an infinitely deep hard-wall trap. 
Numerical calculation is carried out by employing a standard Fourier 
transform, as follows from the Fermi-Bose transformation for a time-dependent 
Lieb-Liniger gas. We study the evolution of the momentum distribution, the 
real-space single-particle density, and the occupancies of natural orbitals.
Our numerical calculation allows us to explore the behavior of these observables 
in the transient regime of the expansion, where they are non-trivially affected 
by the particle interactions. 
We derive analytically (by using the stationary phase approximation) the formula 
which connects the asymptotic shape of the momentum distribution and the initial state. 
For sufficiently large times the momentum distribution coincides (up 
to a simple scaling transformation) with the shape of the real-space 
single-particle density (the expansion is asymptotically ballistic). 
Our analytical and numerical results are in good agreement. 
\end{abstract}

\maketitle

\section{Introduction}
\label{sec:intro}
Exactly solvable models describing interacting bosons 
in one-dimension (1D) have been studied over decades since the pioneering work 
of Girardeau \cite{Girardeau1960}, and Lieb and Liniger \cite{Lieb1963}. 
The interest in these models is greatly stimulated with recent experiments 
\cite{OneD,TG2004,Kinoshita2006}, in which ultracold atomic gases are tightly confined 
in 1D atomic waveguides, such that transverse excitations are suppressed. 
The Lieb-Liniger (LL) model describes 1D bosons with pointlike contact
interactions of a given strength $c$ \cite{Lieb1963}. In the limit of sufficiently 
strong interactions, the LL gas enters the Tonks-Girardeau (TG) regime of 
impenetrable bosons \cite{Girardeau1960}; the TG regime can be obtained at very 
low temperatures, with strong effective interactions, and low linear particle 
densities \cite{Olshanii,Petrov,Dunjko}. An interesting aspect of 1D Bose gases, 
which can be probed experimentally from weakly to the strongly interacting 
regime, is their behavior out of equilibrium (e.g., see Ref. \cite{Kinoshita2006}). 
An exact (analytical or numerical) theoretical calculation of nonequilibrium dynamics 
of a LL gas is a complex many-body problem, which was studied in a few cases 
\cite{Gaudin1983,Girardeau2003,Berman2004,Buljan2008,Jukic2008}. 
A paradigmatic problem in this context is one-dimensional free expansion from an 
initially localized state \cite{Buljan2008,Jukic2008,Ohberg2002,
Rigol2005,Minguzzi2005,DelCampo2006,Gangardt2007}. Quite generally, free expansion 
can be used to provide information on the initial state (e.g., see Refs. 
\cite{Bloch2008,Altman2004, AMRey2008} and references therein).

Free expansion of a LL gas was studied in Ref. \cite{Ohberg2002} by using 
the hydrodynamic approximation \cite{Dunjko}; it was demonstrated that the 
evolution of single-particle (SP) density is not self-similar for 
finite interaction strength $c$. Exact solutions of free expansion 
were studied both in the TG regime ($c=\infty$) \cite{Rigol2005,Minguzzi2005,
DelCampo2006,Gangardt2007}, and for a repulsive LL gas ($0\leq c<\infty$) 
\cite{Buljan2008,Jukic2008}. In the TG regime, it has been shown that the momentum 
distribution approaches that of noninteracting fermions during free expansion 
\cite{Rigol2005,Minguzzi2005}. 
This was shown numerically by using the model of hard-core bosons on the lattice 
in Ref. \cite{Rigol2005}, and by using the stationary phase approximation 
in the continuous TG model (for the initial harmonic confinement) in Ref. 
\cite{Minguzzi2005}. 
For quite general initial conditions, the asymptotic form of the wave functions 
for a freely expanding LL gas was calculated in Ref. \cite{Jukic2008} by using 
the stationary phase approximation (free expansion for a particular family of initial 
conditions was previously considered in \cite{Buljan2008}). 
It was shown that these wave functions vanish at the hyperplanes of contact 
between particles, which is characteristic for TG wave functions 
\cite{Girardeau1960}. 
However, it was emphasized that the properties of such asymptotic states 
can considerably differ from the physical properties of a Tonks-Girardeau gas 
in the ground state of some trapping potential \cite{Jukic2008}
(see also the second item of Ref. \cite{Buljan2008}).
We also point out that expansion dynamics is subject of studies 
in other 1D models; for example particles interacting via an inverse-square 
pair potential \cite{Sutherland2008}, strongly correlated fermions \cite{Heidrich2008}, 
and strongly interacting Bose-Fermi mixtures \cite{Minguzzi2009}. 
To the best of our knowledge, studies of the momentum distribution of an expanding 
LL gas, which are based on exact time-dependent solutions, have not yet been made.

Here we numerically study free expansion of a few Lieb-Liniger bosons, which are initially 
in the ground state of an infinitely deep hard-wall trap. 
The numerical calculation is carried out by employing a standard Fourier 
transform, as follows from the Fermi-Bose transformation for a time-dependent 
Lieb-Liniger gas \cite{Buljan2008,Gaudin1983}. 
We focus on dynamics of one-body observables of the system, in particular the 
momentum distribution, the occupancies of natural orbitals, and also the 
real-space single-particle density. 
Our numerical calculation allows us to explore the behavior of these observables 
in the transient regime of the expansion, where they are non-trivially affected 
by the particle interactions. 
We derive analytically (by using the stationary phase approximation) the formula 
which connects the asymptotic shape of the momentum distribution and the initial state. 
For sufficiently large times the momentum distribution coincides (up 
to a simple scaling transformation) with the shape of the real-space 
single-particle density, reflecting the fact that the expansion is asymptotically 
ballistic. 
The relation between the asymptotic expansion velocity of the LL cloud, 
and the overall energy stored in the system is derived.
Our analytical and numerical results are in good agreement. 

Before proceeding, let us devote a few words to some of the techniques for 
solving LL and TG models. The Bethe ansatz can be used to find the 
eigenstates for LL particles on an infinite line \cite{Lieb1963}, with 
periodic boundary conditions \cite{Lieb1963}, and in an infinitely deep box 
\cite{Gaudin1971}. In the TG limit, both stationary \cite{Girardeau1960} and 
time-dependent \cite{Girardeau2000} wave functions, in an arbitrary 
external potential, are constructed by using the Fermi-Bose mapping, 
i.e., by solving the Schr\" odinger equation for spinless noninteracting fermions, 
after which the fermionic wave function is properly symmetrized to 
describe TG bosons \cite{Girardeau1960,Girardeau2000}. 
In a similar fashion, exact time-dependent LL wave functions, in 
the absence of an external potential and on an infinite line, can be constructed 
by employing the Fermi-Bose mapping operator \cite{Gaudin1983,Buljan2008}, which 
ensures that the so-called cusp-condition (see e.g., \cite{Girardeau2003,Buljan2008}) 
imposed by the interactions is obeyed during time-evolution. 
In the light of the recent experiments \cite{OneD,TG2004,Kinoshita2006}, there has 
been renewed interest in exact studies of LL gases (stationary 
\cite{Busch1998,Muga1998,Sakmann2005,Batchelor2005,Kinezi2006,Sykes2007} and 
time-dependent \cite{Buljan2008,Jukic2008,Girardeau2003}), and time-dependent TG gases 
\cite{Rigol2005,Minguzzi2005,DelCampo2006,Gangardt2007,Girardeau2000,
Girardeau2000a,Busch2003,Rigol2006,Buljan2006,Pezer2007}. 

Even once a wave function describing a LL gas is known, determination of 
its correlation functions and observables such as the momentum distribution is 
difficult \cite{Korepin1993,Jimbo1981,Olshanii2003,Gangardt2003,Astrakharchik2003,
Kheruntsyan2005,Forrester2006,Caux2007}. 
Various methods were developed over the years including the 
Quantum Inverse Scattering Method \cite{Korepin1993,Caux2007}, 
$1/c$ expansions \cite{Jimbo1981} from the TG $(c\rightarrow \infty)$ regime, 
and Quantum Monte Carlo Integration \cite{Astrakharchik2003}. 
In the TG limit, momentum distribution can be analytically 
studied in a few cases (e.g., see \cite{Lenard1964,Forrester2003}). 
Numerical calculation of the TG momentum distribution can be performed 
also for excited and time-dependent states \cite{Rigol2005,Pezer2007};
for hard-core bosons on the lattice see \cite{Rigol2005}; a simple formula 
suitable for numerical calculations was recently derived for the continuous 
TG model \cite{Pezer2007}, and generalized for 1D hard-core anyons \cite{DelCampo2008}.

\section{The Lieb-Liniger model and observables of interest}
\label{sec:LL}
We examine a system of $N$ identical $\delta$-interacting bosons which are 
constrained to one spatial dimension.
The Schr\"odinger equation for that system, in the absence of any external 
potential, is
\begin{equation}
i \frac{\partial \psi_B}{\partial t}=
-\sum_{i=1}^{N}\frac{\partial^2 \psi_B}{\partial x_i^2}+
\sum_{1\leq i < j \leq N} 2c\,\delta(x_i-x_j)\psi_B.
\label{LLmodel}
\end{equation}
The strength of interaction is described by a parameter $c$
(here we consider repulsive interactions $c>0$). 
The initial condition $\psi_B(x_1,x_2,\ldots,x_N,t=0)$ is a localized state,
e.g., the LL ground state in some external trapping potential.
Since we are dealing with symmetric (bosonic) wave functions, it is 
convenient to write Eq. (\ref{LLmodel}) in one permutation sector of the 
configuration space, $R_1:x_1<x_2<\ldots<x_N$,
\begin{equation}
i \frac{\partial \psi_B}{\partial t}=
-\sum_{i=1}^{N}\frac{\partial^2 \psi_B}{\partial x_i^2};
\label{free}
\end{equation}
the $\delta$-function interactions are equivalent to the boundary condition
\cite{Lieb1963},
\begin{equation}
\left [
1-\frac{1}{c}
\left (
\frac{\partial}{\partial x_{j+1}}-\frac{\partial}{\partial x_j}
\right)
\right]_{x_{j+1}=x_j}\psi_B=0.
\label{interactions}
\end{equation}
It is worth to mention that the so-called cusp condition in the TG 
limit (when $c \rightarrow \infty$) reduces to the condition that the wave 
function vanishes whenever any of the two particles touch.

The time-dependent Schr\"odinger equation (\ref{LLmodel}) can be solved exactly
by employing the Fermi-Bose mapping operator \cite{Gaudin1983,Buljan2008},

\begin{equation}
\hat O_c=\prod_{1\leq i < j \leq N}
\left[
\mbox{sgn}(x_j-x_i)+\frac{1}{c}
\left(
\frac{\partial}{\partial x_{j}}-
\frac{\partial}{\partial x_{i}}
\right)
\right].
\label{oO}
\end{equation}
If we find a fully antisymmetric (fermionic) wave function $\psi_F$ which 
obeys 
\begin{equation}
i \frac{\partial \psi_F}{\partial t}=
-\sum_{i=1}^{N}\frac{\partial^2 \psi_F}{\partial x_i^2},
\label{SchF}
\end{equation}
then the wave function
\begin{equation}
\psi_{B,c}= {\mathcal N}_{c} \hat O_c \psi_F,
\label{ansatz}
\end{equation}
where ${\mathcal N}_{c}$ is a normalization constant, obeys Eq. 
(\ref{LLmodel}) \cite{Gaudin1983,Buljan2008,Jukic2008}.

This means that free expansion solutions can be found by using 
Fourier transform. Let
\begin{align}
&\psi_F(x_1,\ldots,x_N,t)
=\int dk_1 \cdots dk_N
\nonumber \\
&\quad\times\ \tilde \psi_F(k_1,\ldots,k_N)
e^{i \sum_{j=1}^{N} [k_j x_j - \omega(k_j) t]},
\label{psiFt}
\end{align}
where $\omega(k)=k^2$, denote an antisymmetric wave function $\psi_F$, 
which evidently obeys Eq. (\ref{SchF}); here 
\begin{align}
&\tilde \psi_F(k_1,\ldots,k_N)
=\frac{1}{(2\pi)^N}
\int dx_1 \cdots dx_N
\nonumber \\
&\quad\times\ \psi_F(x_1,\ldots,x_N,t=0)
e^{-i \sum_{j=1}^{N} k_j x_j}.
\end{align}
is the Fourier transform of $\psi_F$ at $t=0$ \cite{Jukic2008}. 
By acting with the Fermi-Bose mapping operator on $\psi_F$ 
[see Eq. (\ref{ansatz})] we obtain a time-dependent wave function of 
a freely expanding LL gas:
\begin{align}
&\psi_{B,c}(x_1,\ldots,x_N,t)=\int dk_1 \cdots dk_N
\nonumber \\
&\quad\times\ G(k_1,\ldots,k_N)
e^{i \sum_{j=1}^{N} [k_j x_j - k_j^2 t]},
\label{psiBt}
\end{align}
where
\begin{align}
&G(k_1,\ldots,k_N) \equiv {\mathcal N}_{c} \tilde \psi_F(k_1,\ldots,k_N)
\nonumber \\
&\quad\times\
\prod_{1\leq i<j\leq N}
\left[\mbox{sgn}(x_j-x_i)+\frac{i}{c}(k_j-k_i)\right].
\label{gk}
\end{align}
The information on initial conditions is contained in $G(k_1,\ldots,k_N)$
[that is, within $\tilde \psi_F(k_1,\ldots,k_N)$]. 
If we know $G(k_1,\ldots,k_N)$ in $R_1$, by calculating 
the Fourier transform in Eq. (\ref{psiBt}), we find the wave function 
$\psi_{B,c}$ at some finite time $t>0$ in $R_1$. In Ref. \cite{Jukic2008}, 
$\tilde \psi_F(k_1,\ldots,k_N)$ was found to be proportional to the 
projection coefficients $b(k_1,\ldots,k_N)$ of the initial bosonic wave functions 
onto the LL eigenstates in free space. This connection allows us to calculate 
$G(k_1,\ldots,k_N)$ for a few particles, which are in the ground state 
of a LL gas in a box potential \cite{Gaudin1971}, and to study free 
expansion from such an initial state; this is performed in Sec. \ref{sec:Expansion} 
and Appendix \ref{app:init}. 
For clarity, it should be noted that $G(k_1,\ldots,k_N)$ also depends 
on the coordinates $x_j$ through the $\mbox{sgn}(x_j-x_i)$ terms ($G(k_1,\ldots,k_N)$ is 
not the Fourier transform of the bosonic wave function \cite{Jukic2008}). 

In principle, from the time-dependent LL wave function $\psi_{B,c}(x_1,\ldots,x_N,t)$ 
one can extract the physically relevant observables (in practice, this is a difficult
task). Here we consider one-body observables contained within the reduced single-particle 
density matrix (RSPDM), 
\begin{align}
\rho_{B,c} (x,y,t) = 
&N \int dx_2 \cdots dx_N \psi_{B,c}(x,x_2,\ldots,x_N,t)^{*} \nonumber \\
&\times \psi_{B,c}(y,x_2,\ldots,x_N,t).
\label{rhoBC}
\end{align}
The SP density in real space is simply $\rho_{B,c} (x,x,t)$, whereas 
the momentum distribution is defined as 
\begin{equation}
n_B(k,t)=\frac{1}{2 \pi} \int dx dy e^{ik(x-y)} \rho_{B,c} (x,y,t).
\label{nBK}
\end{equation}
The eigenfunctions of the RSPDM, $\Phi_i(x,t)$ are called the natural orbitals (NOs),
\begin{equation}
\int dx \rho_{B,c}(x,y,t) \Phi_i(x,t) = \lambda_i(t) \Phi_i(y,t), \;\;\; i=1,2,\ldots;
\end{equation}
the eigenvalues $\lambda_i(t)$ are the occupancies of these orbitals. 
Apparently, in a nonequilibrium situation, the effective single particle states 
$\Phi_i(x,t)$ and their occupancies $\lambda_i(t)$ may change in time. 

\section{Asymptotic form of the momentum distribution}
\label{sec:AsymptoticMD}

In this section we derive 
the asymptotic form of the momentum distribution of a Lieb-Liniger gas
after free expansion from an initially localized state defined by 
$G(k_1,\ldots,k_N)$ [we should keep in mind that $G(k_1,\ldots,k_N)$ 
also depends upon the coordinates $x_j$ via the $\mbox{sgn}$ functions, 
see Eq. (\ref{gk})].
The momentum distribution defined in Eq. (\ref{nBK}) can be rewritten by using 
Eqs. (\ref{psiBt}) and (\ref{rhoBC}) as
\begin{eqnarray}
&&n_B(k) = \frac{N}{2 \pi} \int dx dy e^{ik(x-y)} \int dx_2 \cdots dx_N \nonumber\\
&\times& \left( \int dk_1 \cdots dk_N G(k_1,\ldots,k_N) 
	e^{i \sum_{j=1}^N (k_j x_j-k_j^2 t)}\right)_{x_1 = x}^{*} \nonumber\\
&\times& \left( \int dq_1 \cdots dq_N G(q_1,\ldots,q_N) 
	e^{i \sum_{j=1}^N (q_j x_j-q_j^2 t)}\right)_{x_1 = y} \nonumber\\
&=& \frac{N}{2 \pi} \int dx_2 \cdots dx_N \int dx dy  
	\, dk_1 \cdots dk_N \, dq_1 \cdots dq_N \nonumber\\
&\times&  G(k_1,\ldots,k_N)^{*}|_{x_1 = x}\ G(q_1,\ldots,q_N)|_{x_1 = y}\ e^{i \phi},
\label{MDapp1}
\end{eqnarray}
where the phase $\phi$ is
\begin{eqnarray}
&&\phi(k_1,\ldots,k_N,q_1,\ldots,q_N,x,y) = - \sum_{j=2}^N(k_j x_j-k_j^2 t) \nonumber\\
&+& \sum_{j=2}^N(q_j x_j-q_j^2 t) -k_1 x+k_1^2 t + q_1 y - q_1^2 t + k(x-y). \nonumber
\end{eqnarray}
The integrals over $k_1,\ldots,k_N,q_1,\ldots,q_N,x,y$ in Eq. (\ref{MDapp1}) 
are evaluated with the stationary phase approximation.
The point of stationary phase is defined by the following equations: 
$$\left.\frac{\partial \phi}{\partial k_j}\right|_{{k^{'}_j}}=
  \left.\frac{\partial \phi}{\partial q_j}\right|_{{q^{'}_j}}=
  \left.\frac{\partial \phi}{\partial x}\right|_{{x^{'}}}=
  \left.\frac{\partial \phi}{\partial y}\right|_{{y^{'}}}=0, \ \mbox{for}\ 1\leq j \leq N.$$
The stationary phase point is:
\begin{align}
& k_j^{'}=q_j^{'}=x_j/(2t),\ \mbox{for}\ 2\leq j \leq N,\nonumber \\ 
& k_1^{'}=q_1^{'}=k,\ \mbox{and} \nonumber \\
& x^{'}=y^{'}=2kt.
\label{stphpt}
\end{align}
The phase $\phi$ can be rewritten as
\begin{align}
\phi  & =  t \sum_{j=2}^{N} 
\left[ (k_j-\frac{x_j}{2t})^2 -(q_j-\frac{x_j}{2t})^2 \right] \nonumber \\
& + [(k-k_1)x+k_1^2 t] - [(k-q_1)y+q_1^2 t].
\end{align}
We notice that $\phi(k_1^{'},\ldots,k_N^{'},q_1^{'},\ldots,q_N^{'},x^{'},y^{'})=0.$
In the stationary phase approximation, the function $G$ in Eq. (\ref{MDapp1})
is evaluated at the stationary phase point defined in Eq. (\ref{stphpt}), 
which yields 
\begin{align}
&n_{B,\infty}(k) \approx \frac{N}{2 \pi} \int dx_2 \cdots dx_N 
\left |G(k,\frac{x_2}{2t},\ldots,\frac{x_N}{2t}) \right|^{2} \nonumber\\
&\times \int dx dk_1 e^{i[(k-k_1)x+k_1^2 t]} \ \int dy dq_1 e^{-i[(k-q_1)y+q_1^2 t]} \nonumber\\
&\times 
\left(\int dk_2 e^{i t (k_2-\frac{x_2}{2t})^2}\right)^{N-1} 
\left(\int dq_2 e^{-i t (q_2-\frac{x_2}{2t})^2}\right)^{N-1} \nonumber\\
&= \frac{N}{(2 \pi)^3} \int dx_2 \cdots dx_N 
\left|G(k,\frac{x_2}{2t},\ldots,\frac{x_N}{2t}) \right|^{2}  \nonumber\\
&\times e^{i k^2 t} e^{-i k^2 t} 
\left(\sqrt{\frac{\pi}{t}} e^{i \pi/4} \right)^{N-1} 
\left(\sqrt{\frac{\pi}{t}} e^{-i \pi/4} \right)^{N-1}.
\label{MDapp2}
\end{align}
It is convenient now to introduce variables $\xi_j=x_j/t$; from 
(\ref{MDapp2}) we obtain the asymptotic form of the momentum 
distribution of a freely expanding LL gas 
\begin{align}
n_{B, \infty}(k) \propto &\int d\xi_2 \cdots d\xi_N 
\ |G(k,\xi_2/2,\ldots,\xi_N/2)|^2.
\label{MDasym}
\end{align}
We note that in the asymptotic regime, the momentum distribution 
acquires the same functional form as the asymptotic SP density.
In the asymptotic regime, the SP density exhibits self-similar 
(ballistic) expansion (this is not true in the transient period preceding the 
asymptotic regime, see Ref. \cite{Ohberg2002}). It is most convenient
to express the asymptotic SP density in variable $\xi=x/t$ 
(see Ref. \cite{Jukic2008}),
\begin{align}
\rho_{\infty}(\xi) \propto &\int d\xi_2 \cdots d\xi_N 
\ |G(\xi/2,\xi_2/2,\ldots,\xi_N/2)|^2;
\label{SPasym}
\end{align}
we normalize $\rho_{\infty}(\xi)$ such that $\int \rho_{\infty}(\xi) d\xi = N$.
The variable $\xi=x/t$ has units of velocity; the self-similar 
asymptotic SP density can be interpreted as the distribution of velocities of 
particles in a gas, which is in a simple manner related to the momentum distribution 
$n_{B, \infty}(k)$.

Equation (\ref{MDasym}) can be thought of as a generalization of the dynamical 
fermionization of the momentum distribution which has been demonstrated for a 
freely expanding TG gas ($c\rightarrow\infty$) \cite{Rigol2005,Minguzzi2005}.
Free expansion in the TG regime is solved by the Fermi-Bose mapping 
\cite{Girardeau1960,Girardeau2000}. In this regime, the SP density is identical 
on both sides of the map. Since fermions are noninteracting, the asymptotic form 
of the SP density (for both TG bosons and free fermions) is identical to the 
fermionic momentum distribution, which does not change in time. 
Equations (\ref{MDasym}) and (\ref{SPasym}) immediately yield that the 
asymptotic momentum distribution for TG bosons has the same shape as the asymptotic 
SP density, which has the shape of the fermionic momentum distribution,
i.e., we obtain the result of Refs. \cite{Rigol2005,Minguzzi2005}.
We also note that equivalent relation between the asymptotic SP density and momentum distribution 
was found in Ref. \cite{Sutherland2008} for a different model with emphasis that the time of
flight measurements do not give the initial momentum distribution.
The derived formula (\ref{MDasym}) is verified numerically on a particular example
in the next section.

\section{Free expansion from a box: 
Dynamics of the momentum distribution and the occupancies $\lambda_i(t)$}
\label{sec:Expansion}

In this section we calculate free expansion of three LL bosons, which are initially 
(at $t=0$) in the ground state in an infinitely deep box of length $L=\pi$. 
The analytical expression for the LL box ground state has been found in Ref. \cite{Gaudin1971}. 
By using this result it is straightforward to calculate $G(k_1,k_2,k_3)$
(which depends on the interaction strength $c$) for this particular initial condition;
we have outlined this calculation in Appendix \ref{app:init} for $N$ particles. 
The next step is calculation of the Fourier integral in Eq. (\ref{psiBt}), which 
is performed numerically by employing the Fast Fourier Transform algorithm. 
From the numerically obtained LL wave function $\psi_{B,c}(x_1,x_2,x_3,t)$ 
we calculate the momentum distribution $n_B(k,t)$, the SP density $\rho_{c}(x,t)$, 
natural orbitals and their occupancies, and study their evolution during free 
expansion from the box ground state. 

\begin{figure}
\begin{center}
\includegraphics[width=0.5 \textwidth ]{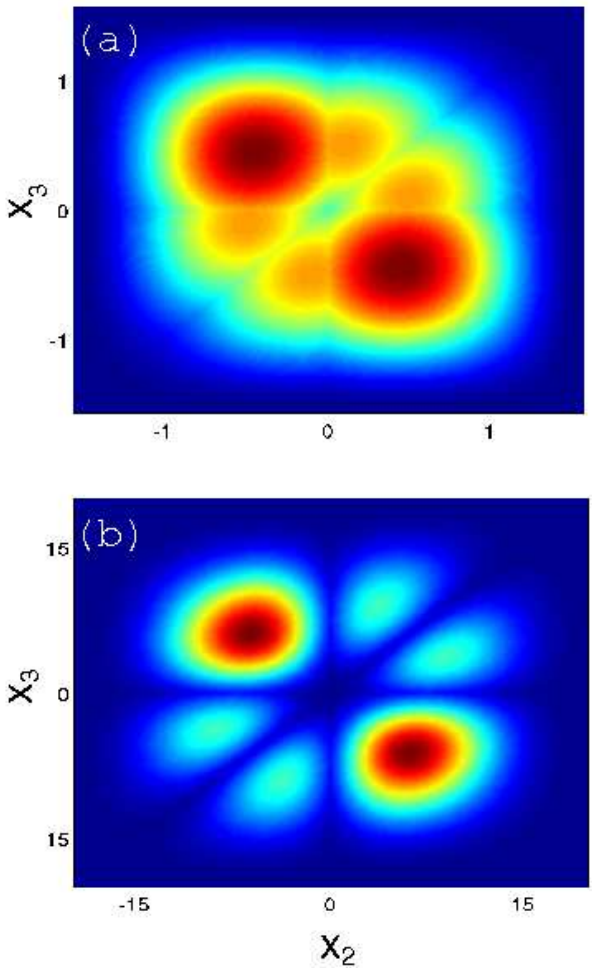}
\caption{ \label{contours}
(color online) Contour plots of $|\psi_{B,c}(0,x_2,x_3,t)|^2$ for $c=1$ at
(a) t=0, and (b) t=3. As the time $t$ increases, the probability density 
at the hyperplanes where particles are in contact decreases.
}
\end{center}
\end{figure}

First let us explore the dynamics of the wave function $\psi_{B,c}(x_1,x_2,x_3,t)$. 
Figure \ref{contours} displays contour plots of the probability density 
$|\psi_{B,c}(0,x_2,x_3,t)|^2$ for $c=1$, at two different times, $t=0$ and $t=3$. 
We see that as the LL gas expands, the probability density decreases at the 
hyperplanes $x_i=x_j$ ($i\neq j$) where the particles are in contact. This is in 
agreement with the result of Ref. \cite{Jukic2008}, where it was shown (by using 
the stationary phase approximation) that the leading term of 
$\psi_{B,c}(\xi_1 t,\xi_2 t,\xi_3 t,t)$ has Tonks-Girardeau form for sufficiently 
large $t$; that is, the leading term is zero for $\xi_i=\xi_j$ ($i\neq j$). 
However, this does not necessarily mean that the properties of such an asymptotic state 
correspond to the properties of a TG gas, which was usually studied in the 
ground state of some external potential. 
For example, suppose that the initial state is a weakly correlated 
ground state in the box; despite the fact that, during expansion, 
the particles get strongly correlated in the close vicinity of the hyperplanes 
of contact, the absence of correlations in the initial state survives as an overall 
feature through to the asymptotic state 
(see the discussion in Ref. \cite{Jukic2008} and the second item of Ref. \cite{Buljan2008}).
Thus, even though that the asymptotic state is described by a wave function 
with the TG structure, the physical properties of the expanded gas can considerably 
differ from the properties of a TG gas. 

%

\begin{figure}
\begin{center}
\includegraphics[width=0.45 \textwidth ]{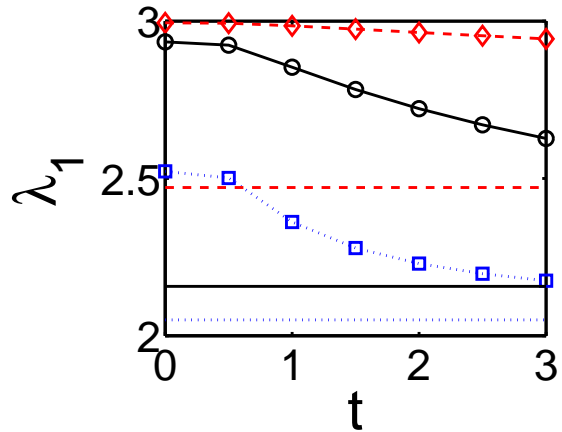}
\caption{ \label{naturals}
(color online) The lowest natural orbital $\lambda_1(t)$ as a function of time for 
three values of $c$. Red diamonds (dashed line) is for $c=0.25$, black
circles (solid line) for $c=1$, and blue squares (dotted line) for $c=5$;
the lines connecting the markers are guides for the eye. The corresponding 
horizontal lines without markers denote the asymptotic occupancies, 
calculated from the asymptotic wave functions \cite{Jukic2008}.
}
\end{center}
\end{figure}

In order to further study the properties of the state in expansion, 
Fig. \ref{naturals} illustrates the occupation of the lowest natural orbital 
in time, $\lambda_1(t)$, for several values of $c$.
The asymptotic values of the occupancies, which are obtained by using 
the asymptotic forms of the wave functions \cite{Jukic2008}, are indicated
with horizontal lines. We observe that the occupancy of the leading NO, 
$\lambda_1(t)$, decreases during time evolution. However, 
for the plotted interaction strengths, the decrease of $\lambda_1(t)$
is not too large. This means that the coherence of the system 
(described by the occupations of the natural orbitals) for the plotted 
parameters only partially decreases during free expansion due to the interactions. 
It should be noted that in the TG limit $c\rightarrow\infty$, 
for hard-core bosons on the lattice \cite{Rigol2005}, 
it has been shown that the leading natural orbitals slightly increase
during free expansion \cite{Rigol2005}, which differs from the finite 
$c$ results obtained here. It is reasonable to associate 
the decrease of $\lambda_1(t)$ to the change of the LL wave
functions at the hyperplanes of contact; this change does not occur in the
TG regime, where the wave functions are zero at the contact hyperplanes at
any time of the expansion.

%

\begin{figure}
\begin{center}
\includegraphics[width=0.4 \textwidth ]{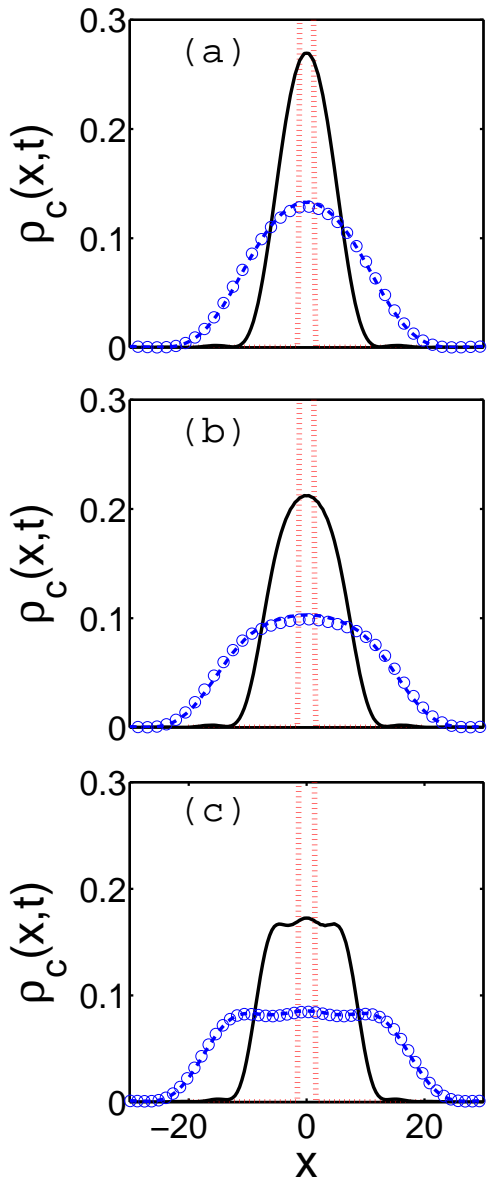}
\caption{ \label{timeevolRHO}
(color online) Evolution of the $x$-space density in time for various interaction
strengths $c$: (a) $c=0.25$, 
at $t=0$ (red dotted line), 
$t=2$ (solid black line), 
$t=4$ (blue dashed line);
(b) $c=1$, 
at $t=0$ (red dotted line), 
$t=2$ (solid black line), 
$t=4$ (blue dashed line);
(c) $c=10$, 
at $t=0$ (red dotted line), 
$t=1$ (solid black line), 
$t=3$ (blue dashed line).
The asymptotic $x$-space density $\rho_{\infty}(\xi)$ (circles), 
is plotted as a function of $x= \xi t$ corresponding to the largest time
in each subplot.
}
\end{center}
\end{figure}

\begin{figure}
\begin{center}
\includegraphics[width=0.415 \textwidth ]{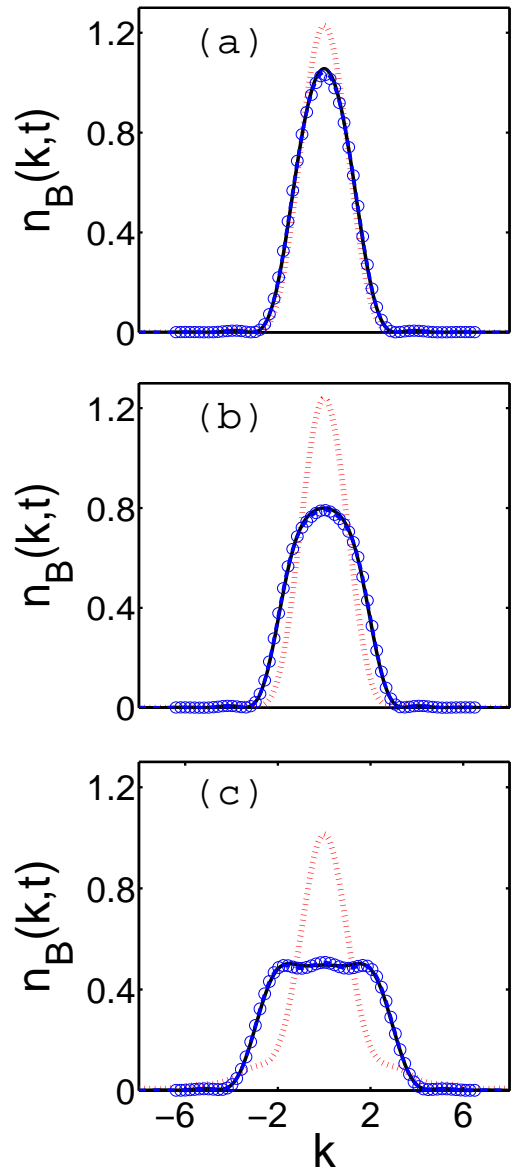}
\caption{ \label{timeevolNK}
(color online) Evolution of the momentum distribution in time for various interaction
strengths $c$. The lines and colors for different values of $c$ and $t$ are identical as 
in Fig \ref{timeevolRHO}. Solid black and blue dashed line are almost indistinguishable.
}
\end{center}
\end{figure}

Let us explore the dynamics of the momentum distribution $n_{B}(k,t)$, 
and its connection to the SP density $\rho_{c}(x,t)$ at large times $t$. 
The time-evolution of $\rho_{c}(x,t)$ and $n_{B}(k,t)$ is illustrated in Figs. 
\ref{timeevolRHO} and \ref{timeevolNK}; we display $x$- and $k$-space 
densities for various values of the parameter $c$, at several times $t$.  
Initially, all momentum distributions have a typical bosonic property: 
they peak at $k=0$. We observe that the qualitative changes in the shape of 
$n_{B}(k,t)$ are more pronounced for larger values of $c$. 
Circles in Figs. \ref{timeevolRHO} and \ref{timeevolNK} show 
the asymptotic values calculated by using Eqs. (\ref{MDasym}) and (\ref{SPasym}). 
We see that at the maximal value of time $t$ in the plots, the momentum 
distribution agrees well with that obtained with the stationary phase approximation 
in Eq. (\ref{MDasym}). 
Our numerical calculation is in agreement with the findings presented in 
Eqs. (\ref{MDasym}) and (\ref{SPasym}). 
We would like to point out that, even though the observables $n_{B}(k,t)$ 
and $\rho_{c}(x,t)$ are well approximated by the stationary phase 
approximation at the maximal expansion time reached in our numerical 
simulations (see Figs. \ref{timeevolRHO} and \ref{timeevolNK}), 
the system is strictly speaking not yet fully in the asymptotic regime
(e.g., note that the occupancies of the natural orbitals have not reached their 
asymptotic values) and even better agreement should be expected at larger times.
Unfortunately, the maximal time allowed in our numerical calculations 
is limited by the computer memory and time. 
%
%
%

\begin{figure}
\begin{center}
\includegraphics[width=0.45 \textwidth ]{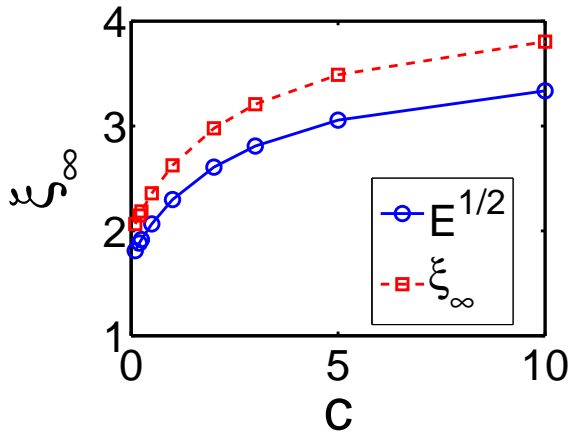}
\caption{ \label{velocity}
(color online) Asymptotic expansion velocity, $\xi_{\infty}$ (squares, 
dashed line), and the square root of the total energy, $\sqrt{E}$ (circles, 
solid line) for various interaction strengths $c$; 
lines serve to guide the eye (see text for details).
}
\end{center}
\end{figure}

In order to further study the 
asymptotic forms of the momentum distribution and the SP density, 
let us calculate the asymptotic expansion velocity as a function of 
the interaction parameter $c$. Since different parts of the cloud expand at different 
velocities, a definition of this quantity has a certain degree of freedom.
Here we define this quantity as a root mean square of the asymptotic SP density 
\cite{FWHMref} in variable $\xi=x/t$ (i.e., velocity):
\begin{equation}
\xi_{\infty} =
\sqrt{\frac{1}{N} \int \xi^2 \rho_{\infty}(\xi) d\xi};
\end{equation}
the factor $1/N$ simply reflects the fact that 
$\rho_{\infty}(\xi)$ is normalized to the number of particles $N$.
The asymptotic velocity $\xi_{\infty}$ is connected to the total energy $E$ stored 
in the system. 
During free expansion, the interaction energy is transferred to the 
kinetic energy; in the asymptotic regime all of the energy is kinetic,
and it can be expressed via the momentum distribution:
\begin{equation}
E =\int k^2 n_{B,\infty}(k) dk.
\label{KineticEnergy}
\end{equation}
By using Eqs. (\ref{MDasym}) and (\ref{SPasym}), we obtain
\begin{equation}
\xi_{\infty} = \sqrt{\frac{4}{N}} \sqrt{E},
\label{relation_vE}
\end{equation}
that is, $E=N \xi_{\infty}^2/4$ which is the classical expression for the 
kinetic energy of $N$ particles with velocity $\xi_{\infty}$ and mass $1/2$
(recall that we use units where the kinetic energy operator in Eq. (\ref{LLmodel})
is $-\sum_{i=1}^{N} \partial^2 / \partial x_i^2$). 
The quantities $\sqrt{E}$ and $\xi_{\infty}$ are displayed in Fig. \ref{velocity}
for various values of the interaction strength $c$; the plots 
underpin Eq. (\ref{relation_vE}). The total energy was calculated simply as 
$E=q_1^2+q_2^2+q_3^2$ where quasimomenta $q_i$ are obtained by solving 
transcendental Bethe equations for the initial state \cite{Gaudin1971}
(see Appendix \ref{app:init}). 
The asymptotic velocity was obtained via Eq. (\ref{KineticEnergy}) by 
numerical integration. 
Our numerical calculations are in good agreement (better than $99\%$) 
with Eq. (\ref{relation_vE}); we attribute the discrepancy to inaccuracy 
of the numerical integration. 

%
%

\section{Conclusion}
\label{sec:concl}

We have numerically studied free expansion of a few Lieb-Liniger bosons, which are initially 
in the ground state of an infinitely deep hard-wall trap. 
This numerical calculation has been carried out by employing a standard Fourier 
transform, as follows from the Fermi-Bose transformation for a time-dependent 
Lieb-Liniger gas. 
We have studied the evolution of the momentum distribution, the real-space single-particle 
density,  and the occupancies of natural orbitals, both in the non-trivial 
transient regime of the expansion and asymptotically.  
We have derived analytically (by using the stationary phase approximation) the formula 
which connects the asymptotic shape of the momentum distribution and the initial state. 
For sufficiently large times the momentum distribution coincides (up 
to a scaling transformation) with the shape of the real-space 
single-particle density (the expansion is asymptotically ballistic). 
This result can be considered as a generalization 
of the dynamical fermionization of the momentum distribution in the Tonks-Girardeau 
regime, which has been pointed to occur in the course of free expansion 
\cite{Rigol2005,Minguzzi2005}. 
We have shown that the occupancy of the lowest natural orbital of the system 
decreases with time while approaching its asymptotic value. 
This was related to the build-up of correlations of the hyperplanes of 
contact of the particles.
Finally, we have calculated the expansion velocity in asymptotic regime and 
pointed out its relation to the overall energy of the system.

In order to gain further understanding of a freely expanding LL gas, 
it would be desirable to investigate transient dynamics of the observables
for larger number of particles, and also for different initial conditions
(e.g., the ground state of a LL gas in different initial trapping potentials).


\acknowledgments
We acknowledge useful discussions with Adolfo del Campo, 
Thomas Gasenzer, Anna Minguzzi, Robert Pezer and Marcos Rigol.
This work is supported by the Croatian Ministry of Science, 
Grant No. 119-0000000-1015.

\begin{appendix}
\section{The function $G(k_1,\ldots,k_N)$ for the box ground state}
\label{app:init}

In Sec. \ref{sec:Expansion} we have studied free expansion of three LL bosons, 
which are initially (at $t=0$) in the ground state in an infinitely deep box 
of length $L=\pi$.
Here we present exact analytical expression for function
$G(\{ k \})\equiv G(k_1,\ldots,k_N)$ for this particular case.
First, we use the connection between $\tilde \psi_F(k_1,\ldots,k_N)$ 
and the projection coefficients $b(k_1,\ldots,k_N)$ of the initial bosonic wave 
functions onto the LL eigenstates in free space \cite{Jukic2008} to rewrite
the expression for $G$:
\begin{align}
G(\{ k \}) &= N! \mathcal{N}(\{ k \}) b(\{ k \}) \nonumber \\
&\times\
\prod_{1\leq i<j\leq N}
[\mbox{sgn}(x_j-x_i)+\frac{i}{c}(k_j-k_i)].
\label{connection1} 
\end{align}
Here, $\mathcal{N}(\{ k \})$ is the normalization constant for LL eigenstates 
in free space \cite{Korepin1993},
\begin{equation}
\frac{1}{\mathcal{N}({\{ k \}})}=
\sqrt{ (2 \pi)^N N! \prod_{i<j}{\left[1+\left(\frac{k_j-k_i}{c}\right)^2 \right]}},
\end{equation}
and coefficients $b(\{ k \})$ are found by using the solution 
for the LL box ground state \cite{Gaudin1971},
\begin{align}
&b(\{ k \}) \propto \mathcal{N}(\{ k \}) \sum_{P^{'}} (-1)^{P^{'}} \prod_{1\leq i<j\leq N}
\Big[1-\frac{i}{c}(k_{P^{'}j}-k_{P^{'}i})\Big] \nonumber\\
&\times \sum_{\{ \epsilon \}} \sum_{P} \epsilon_1 \cdots \epsilon_N \prod_{1\leq i<j\leq N}
(1-\frac{ic}{q_i+q_j}) (1+\frac{ic}{q_{Pi}-q_{Pj}})\nonumber\\
&\times  \int_{-L/2}^{L/2} dx_1 \int_{x_1}^{L/2} dx_2 \cdots \int_{x_{N-1}}^{L/2} dx_N \nonumber\\
&\times \exp{\left\{i \sum_{j=1}^{N} \left[\left(q_{Pj}-k_{P^{'}j}\right)x_j-q_{Pj}\frac{L}{2}\right]\right\}}.  
\label{projections}
\end{align}
In the expression above, summations are taken over all permutations $P$ and $P^{'}$ 
which are of order $N$, whereas the set $\{\epsilon\}$ is defined such that each 
$\epsilon_i$ is either $+1$ or $-1$ (here $i=1,\ldots,N$, i.e., there are $2^N$ 
combinations in the set $\{\epsilon\}$). The ground state quasimomenta are defined as 
$q_i=\epsilon_i |q_i|$, for $i=1,\ldots,N$, and their magnitudes $|q_i|$ are found 
by solving (numerically) the system of coupled transcendental equations \cite{Gaudin1971}
\begin{equation}
|q_i| L = \pi + \sum_{j \neq i} \left( \tan^{-1} \frac{c}{|q_i|-|q_j|} + 
\tan^{-1} \frac{c}{|q_i|+|q_j|}\right).
\label{transc}
\end{equation}
Finally, let us mention that the constant of proportionality in Eq. 
(\ref{projections}) is fixed such that the wave 
function $\psi_{B,c}$ is properly normalized.

\end{appendix}



\begin{thebibliography}{99}
\bibitem{FWHMref}
An alternate definition could be to define the 
asymptotic velocity as the half width at half maximum of the 
distribution $N^{-1} \rho_{\infty}(\xi)$.
\end{thebibliography}
\end{document}